\documentclass[twocolumn,showpacs,preprintnumbers,amsmath,amssymb]{revtex4}
\usepackage{graphicx}% Include figure files
\usepackage{bm}% bold math

\def\eq#1{Eq.\,(\ref{#1})}
\begin{document}

\preprint{APS/123-QED}

\title{A Case of Almost Redundant Components
in the Three-Alpha Faddeev Equations}
% Force line breaks with \\

\author{Y. Fujiwara}
\affiliation{Department of Physics, Kyoto University, 
Kyoto 606-8502, Japan}%
 \email{fujiwara@ruby.scphys.kyoto-u.ac.jp}
%Lines break automatically or can be forced with \\
\author{Y. Suzuki}%
\affiliation{Department of Physics, Niigata University,
Niigata 950-2181, Japan}

\author{M. Kohno}
\affiliation{Physics Division, Kyushu Dental College,
Kitakyushu 803-8580, Japan}

\date{\today}% It is always \today, today,
             %  but any date may be explicitly specified

\begin{abstract}
The three-alpha orthogonality condition model using the
Pauli-forbidden bound states of the Buck, Friedlich
and Wheatly $\alpha \alpha$ potential can yield
a compact three-alpha ground state with a large binding energy,
in which a small admixture of the redundant components
can never be eliminated.
\end{abstract}

% PACS, the Physics and Astronomy
                             % Classification Scheme.
%\keywords{Suggested keywords}%Use showkeys class option if keyword
                              %display desired
\maketitle

%\end{document}

%\widetext

As a typical example of quantum-mechanical three-body systems,
the three-alpha-particle ($3\alpha$) model
for $\hbox{}^{12}\hbox{C}$ has
been extensively studied from various viewpoints.\,\cite{SU68}
It is well known that the microscopic structure
of the $\alpha$-cluster plays an important role to create
damped inner oscillations in the relative wave functions
of the two $\alpha$-clusters, which can be described
phenomenologically as the short-range repulsive core   
in the $2\alpha$ system. In the $3\alpha$ system, the amplitudes of
these inner oscillations can be enhanced
by the existence of the third $\alpha$-cluster,
resulting in the formation of the compact
shell-model like ground state of $\hbox{}^{12}\hbox{C}$.\,\cite{HO75}
On the other hand, the loosely bound nature of the $2\alpha$ pair
is still preserved in the excited $0^+$ state at $E_x=7.65$ MeV,
for which much interest is recently paid as a possible candidate
of the $\alpha$ condensation state.\,\cite{TO01,SU02}
This structure change of the $3\alpha$ system can most easily
be simulated in the orthogonality condition model (OCM),
first proposed by Saito \cite{SA68}. 

In a previous publication \cite{ocm03}, we have discussed
a new type of the Faddeev formalism
for the $3\alpha$ system,
in which pairwise $\alpha$-clusters interact via the
Buck, Friedlich and Wheatly potential (BFW potential) \cite{BF77}. 
In this model, the Pauli-forbidden states
between the two $\alpha$-clusters are
composed of the lowest two $S$-wave bound states
and one $D$-wave bound state of the BFW potential.   
We found that the $3\alpha$ ground-state energy 
is $-19.897$ MeV for this potential, which is contradictory
to the very small binding energy,
$E_{3\alpha}=-0.26$ MeV, from the variational calculations
carried out by Tursunov, Baye, Descouvemont
and Daniel in Refs.\,\cite{TB03,DE03}.
The same situation also happens when we neglect the
Coulomb force between $\alpha$-clusters. Namely, we have
obtained $E_{3\alpha}=-27.748~\hbox{MeV}$,
while theirs $-6.003~\hbox{MeV}$ \cite{tur03}.
These authors comment that our result for the $3\alpha$ ground-state
solution, using the BFW bound-state Pauli-forbidden states,
does not completely eliminate the Pauli-forbidden components. 
Unlike their work our Faddeev solution
contains a small admixture of the redundant components.
Suppose $\Psi=\varphi_\alpha+\varphi_\beta+\varphi_\gamma$ be the
total wave function of the $3\alpha$ system, composed of the
three Faddeev components $\varphi_\alpha$,
$\varphi_\beta$ and $\varphi_\gamma$.
If one sets $f_u \equiv \langle u | \Psi \rangle$ with $u$ being
one of the Pauli-forbidden bound-state solutions,
$N_R=\sum_u \langle f_u|f_u\rangle$ for all three Pauli-forbidden
states is only $(2.6 \sim 2.7) \times 10^{-4}$.
This is a big contrast to the result
for the harmonic oscillator (h.o.) Pauli-forbidden
states $|u \rangle$, since in this case $N_R \sim 10^{-12}$ \cite{ocm03}.
The purpose of this brief report is to show that, if one wants
to keep the shell-model like compact $3\alpha$ ground state,
one cannot help but allowing a small admixture of the
redundant components. In other words, it is impossible
to eliminate this small admixture in the present framework
without giving up the solution with the dominant
shell-model like [3](04) component with the total h.o. quanta $N=8$.
This implies that their solution corresponds
to our second (excited-state) solution.
The energy of this second solution is about $-6$ MeV,
and it has a small (04) component and $N_R \sim 10^{-6}$.

A main problem arises from the second [21]-symmetric component
in the $3\alpha$ Faddeev equation,
which now becomes no longer an exact redundant component,
but ``an almost redundant component'' of the Faddeev equation.
Here we use the Faddeev terminology and the notations used in our
previous publications, Refs.\,\cite{RED} and \cite{ocm03},
but the same analysis is also possible in the various variational
approaches.
In Ref.\,\cite{RED}, we first solve the eigenvalue equation
of the rearrangement matrix
\begin{equation}
\langle u|S|uf^\tau\rangle=\tau |f^\tau\rangle\ ,
\label{eq1}
\end{equation}
where $S=(123)+(123)^2$ and $|f^\tau \rangle$ is normalized
as $\langle f^\tau | f^{\tau^\prime}\rangle
=\delta_{\tau, \tau^\prime}$.
The solution $|f^\tau\rangle$ with $\tau=-1$ gives 
a [21]-symmetric redundant solution $\varphi_\tau=G_0|uf^\tau\rangle$ of
the Faddeev equation, where $G_0$ is the free 3-body Green function.
The Faddeev component $\varphi_\tau$ trivially satisfies
\begin{equation}
\lambda(E)\,\varphi= G_0 \widetilde{T} S \varphi
\quad \hbox{with} \quad \lambda(E)=1\ ,
\label{eq2}
\end{equation}
due to the the orthogonality property, $\widetilde{T}G_0|u\rangle=-|u\rangle$,
of the redundancy-free $\widetilde{T}$-matrix and
the commutability $G_0 S=S G_0$.
For this reason we add an extra term
as in \eq{eq9} below, and determine the bound-state
energy $E$ with $\lambda(E)=1$.
After $E$ is determined in this way,
we again solve the Faddeev equation \eq{eq2} without
this second term.
Then we get three $\lambda(E)=1$ solutions;
one is a real solution and the others are the dual complex
solutions having $\Re e \{\lambda(E)\} \sim 1$ with a small
imaginary part of the order of $10^{-2}$.
The appearance of the complex eigenvalues having opposite signs
in the imaginary part is not excluded since
we are working with an eigenvalue problem of the non-symmetric
kernel, $G_0 \tilde{T} S$. These three solutions are characterized
by the following three $SU_3$ components; $[21]2(20)$,
$[21]4(40)$, and $[3]8(04)$, in the notation $|[f]N(\lambda \mu)
\rangle$ or $[f](\lambda \mu)$ with $N=\lambda+2\mu$.
However, this classification is for the Faddeev component $\varphi$.
If we make $\Psi \sim (1+S)\varphi$, all of these three $\Psi$'s
become [3]-symmetric total wave functions for the $3\alpha$ system,
as long as they are non-zero.
If $|u\rangle$ is the h.o. $(0s)$, $(1s)$ and $(0d)$ states,
the first two [21](20) and [21](40) states
exactly vanish by the $(1+S)$ operation,
which means that these are trivial solutions
of \eq{eq2} with $\lambda(E)=1$.
However, if we use the bound-state $|u\rangle$ of the BFW potential,
the latter [21](40) state becomes almost redundant.
(The same situation also happens for the first [21](20) state,
but the residual component after the $(1+S)$ operation
is very small and less than $10^{-5}$.)
In such a case, we can construct the normalized state
\begin{equation}
\phi^{[3]}_\tau={(1+S)|uf^\tau\rangle \over \sqrt{3(1+\tau)}}\ .
\label{eq3}
\end{equation}
(Note that $(1+S)^2=3(1+S)$.) This becomes [3](04) dominant state.
This can be confirmed by expanding $|uf^\tau\rangle$ in the h.o. basis
and calculating the overlap
of $\phi^{[3]}_\tau$ with the shell-model state,
$|[3]8(04)\rangle$, in the $3\alpha$-cluster representation \cite{RED}.
Here we use a rather compact $\alpha$-cluster
with the h.o. width parameter $\nu=0.28125~\hbox{fm}^{-2}$.
In Table \ref{table1}, the $\phi^{[3]}_\tau$ state generated
from the second [21]4(40) dominant solution $|uf^\tau \rangle$
with $\tau=-0.999037$ involves the [3]8(04) component
with the amplitude 0.865401.
This overlap is obtained from the third overlap
in Table \ref{table1} through
\begin{eqnarray}
\langle [3]8(04)|\phi^{[3]}_\tau\rangle
& = & {\langle [3]8(04)|(1+S)|uf^\tau\rangle \over
\sqrt{3(1+\tau)}} \nonumber \\
& = & \sqrt{{3 \over 1+\tau}} \langle [3]8(04)|uf^\tau\rangle\ ,
\label{eq4}
\end{eqnarray}
since $S^\dagger=S$. The normalization factor $1/\sqrt{3(1+\tau)}$
is the reason of this large overlap.
This immediately reminds us our old experience of the almost
forbidden state \cite{SA73} in 2-cluster systems.
In that case, the almost forbidden state is the cluster excited
state, but in the present case it is the real [3](04) state,
which is generated by the $3\alpha$ symmetrization
from the almost redundant solution of \eq{eq1},
with the dominant [21]-symmetric configuration $|[21]4(40)\rangle$.

It is interesting to note that the transition
from $N=4$ to 8 takes place, since [3](04) is the only
Pauli-allowed state with the lowest h.o. quanta $N=8$.
The reason for this transition is naturally understood if we
recall how we construct the Pauli-forbidden states
in the pairwise OCM for the $3\alpha$ system.
Let us assume for the time being that $|u\rangle$ is
the h.o. Pauli-forbidden states.
We first enumerate the translationally
invariant [3]-symmetric h.o. states
by the Moshinsky rule\,\cite{MO96}.
The elimination of the Pauli-forbidden state
by the diagonalization procedure for the
projection operator $P=\sum_\alpha |u_\alpha \rangle
\langle u_\alpha|$ gives that the lowest Pauli-allowed
state of the $3\alpha$ system is only (04) for $N=8$,
and (62) and (24) for $N=10$, etc.\,\cite{TRGM}
On the other hand, the construction
of the $3\alpha$ Pauli-forbidden states in \eq{eq1} is
exactly equivalent to this elimination procedure
of the Pauli-forbidden state,
as long as the [3]-symmetric basis states are concerned. 
Since the [3]-symmetric Pauli-forbidden states are already
enumerated by \eq{eq3} for the solutions of \eq{eq1} with $\tau > -1$
in the h.o. limit,
$\phi^{[3]}_\tau$ with $\tau \sim -1$ should be an extra state
which is orthogonal to all of these [3]-symmetric Pauli-forbidden states.
Therefore, $\phi^{[3]}_\tau$ in \eq{eq3} generated from the
small deviation from the pure h.o. limit should be the
[3]-symmetric {\em allowed} state with the smallest number
of oscillations, namely, $N=8$ (04) state.  
 
\begin{table}[b]
\caption{Some important overlap amplitudes of the lowest three
solutions of \protect\eq{eq1} with the shell-model states,
when the bound-state $|u\rangle$ of the BFW potential are used
for the $2\alpha$ Pauli-forbidden states.
The h.o. width parameter $\nu=0.28125~\hbox{fm}^{-2}$ is used
for the shell-model wave functions.
}
\label{table1}
\begin{center}
\renewcommand{\arraystretch}{1.4}
\setlength{\tabcolsep}{2.5mm}
\begin{tabular}{crrr}
\hline
\hline
$\tau$ &  $-1.00000$ & $-0.999037$ & $-0.099510$ \\
\hline
$\langle [21]2(20)|uf^\tau\rangle$
& 0.985929 & 0.098412 & 0.000873 \\
$\langle [21]4(40)|uf^\tau\rangle$ 
& 0.000764 & 0.593663 & 0.024228 \\ 
$\langle [3]8(04)|uf^\tau\rangle$
& 0.000657 & 0.015506 & 0.001921 \\
$\langle [3]8(04)|\phi^{[3]}_\tau\rangle$
&   $-$    & 0.865401 & 0.003506 \\
\hline
\hline
\end{tabular}
\end{center}
\end{table}

To be more specific, let us expand the Faddeev
component $\varphi$ by the following basis states:
\begin{enumerate}
\item[1)] [21]-symmetric basis: $\phi^{[21]}_{-1}=|uf^{-1}\rangle$
and the other orthonormalized basis $\phi^{[21]}_\alpha$.
\item[2)] [3]-symmetric basis: $\phi^{[3]}_\tau$ with
$\tau > -1$, given in \eq{eq3}, and the other orthonormalized
basis $\phi^{[3]}_\beta$.
Here it is important to construct these as
\begin{equation}
\langle \phi^{[3]}_\tau|\phi^{[3]}_\beta\rangle=0
\quad \hbox{for} \quad \forall~\tau > -1~\hbox{and}~\beta\ .
\label{eq5}
\end{equation}
\end{enumerate}
We expand $\varphi$ in \eq{eq2} with $\lambda(E)=1$ as
\begin{eqnarray}
\varphi & = & C^{[21]}_{-1}~|uf^{-1}\rangle
+\sum_\alpha C^{[21]}_\alpha~\phi^{[21]}_\alpha \nonumber \\
& & +\sum_{\tau > -1} C^{[3]}_\tau~\phi^{[3]}_\tau
    +\sum_\beta C^{[3]}_\beta~\phi^{[3]}_\beta\ ,
\label{eq6}
\end{eqnarray}
and multiply the resultant equation by $\langle u|$ from the left.
Then, because of the basic relationship,
$\langle u|G_0 \widetilde{T}=-\langle u|$,
the [21]-symmetric part vanishes
by $\langle u|(1+S)|[21]\rangle=0$, and we obtain
\begin{eqnarray}
0 & = & \langle u|\Psi\rangle=\langle u|(1+S)|\varphi\rangle
\nonumber \\
& = & \sum_{\tau > -1} C^{[3]}_\tau \sqrt{3(1+\tau)} |f^\tau\rangle
+\sum_\beta 3 C^{[3]}_\beta \langle u|\phi^{[3]}_\beta\rangle.
\label{eq7}
\end{eqnarray}
Here we further take the matrix element with
some particular $\langle f^\tau|$ with $\tau > -1$.
Then the basis construction in \eq{eq5} gives that the last
term of \eq{eq7} disappears and we are left with
\begin{equation}
C^{[3]}_\tau \sqrt{3(1+\tau)}=0 
\quad \hbox{for} \quad \forall~\tau > -1\ .
\label{eq8}
\end{equation}
This implies that the exact solution
of \eq{eq2} with $\lambda(E)=1$ should not
contain any of the $\phi^{[3]}_\tau$ components with $\tau > -1$;
namely, the $3\alpha$ Pauli-forbidden components.
However, this is correct only within the accuracy
of numerical calculations. For the solution
with $\tau$ far apart from $-1$, $C^{[3]}_\tau \sim 0$ is
certainly true. But, for the second solution of Table \ref{table1}
with $\tau=-0.999037$, $C^{[3]}_\tau$ could be appreciably large,
since $C^{[3]}_\tau \times 0.054 =0$.
(Note that the imaginary part of the dual complex eigenvalues
for the Faddeev equation is of the order of $10^{-2}$.)
In fact, we have a good reason to believe that our ground-state
solution has a dominant $\phi^{[3]}_\tau$ component
with $\tau=-0.999037$, since both of them have
a large [3](04) component.
In a practical calculation, we can classify
this $|uf^{\tau}\rangle$ solution to the complete redundant state
with $\tau=-1$ and solve a ``modified'' Faddeev equation
\begin{eqnarray}
& & \lambda(E)\,\varphi=G_0 \left[\widetilde{T}S
-\sum_{\tau \sim -1}|u f^\tau\rangle
{1 \over \langle u f^\tau|G_0|u f^\tau\rangle}
\langle u f^\tau|\right] \varphi. \nonumber \\
\label{eq9}
\end{eqnarray}
(Otherwise, we obtain unstable complex solutions and the energy
with $\lambda(E)=1$ is not precisely determined.)
The $3\alpha$ ground-state energy obtained by this prescription
is $E_{3\alpha}=-27.625~\hbox{MeV}$, which is very close to
the exact value $-27.748~\hbox{MeV}$ obtained by solving
an improved equation \eq{eq14} or \eq{eq22}.
In this case, the relationship in \eq{eq8} is modified to
\begin{eqnarray}
\langle  uf^\tau |\Psi \rangle
& = & \langle  uf^\tau |1+S|\varphi \rangle \nonumber \\
& = & C^{[3]}_\tau \sqrt{3(1+\tau)}
=-\langle uf^\tau |\varphi \rangle\ .
\label{eq10}
\end{eqnarray}
For the normalized $\varphi$ with $3\langle \varphi|1+S|
\varphi \rangle=1$, the last matrix element of \eq{eq10}
for the ground state is found to be $0.1604 \times 10^{-1}$.
This leads to the value $C^{[3]}_\tau=-0.2984$, which yields
the amplitude of the $\phi^{[3]}_\tau$ component
contained in the total wave function $\Psi$ as
\begin{equation}
\langle \phi^{[3]}_\tau|\Psi\rangle
= \sqrt{{3 \over 1+\tau}} \langle  uf^\tau |\Psi \rangle
= 3 C^{[3]}_\tau=-0.8952\ .
\label{eq11}
\end{equation}
If we assume $\Psi \sim 3 C^{[3]}_\tau \phi^{[3]}_\tau$,
we can approximate the redundant amplitudes as
\begin{equation}
|f_u\rangle = \langle u|\Psi\rangle
\sim 3 C^{[3]}_\tau \langle u| \phi^{[3]}_\tau \rangle
=C^{[3]}_\tau \sqrt{3(1+\tau)} |f^\tau\rangle\ ,
\label{eq12}
\end{equation}
and the redundant component admixed in the ground state is
given by
\begin{equation}
\langle f_u|f_u\rangle \sim {C^{[3]}_\tau}^2
3(1+\tau) =\langle uf^\tau |
\varphi \rangle^2 = 0.26 \times 10^{-3}\ ,
\label{eq13}
\end{equation}
which agrees very well with the
number $(2.6 \sim 2.7)\times 10^{-4}$,
obtained by solving \eq{eq22}.

From the definition of $\phi^{[3]}_\beta$ in \eq{eq5},
it is apparent that none of the $\phi^{[3]}_\beta$ has
the large [3](04) component.
Therefore, if one rejects the second $\phi^{[3]}_\tau$
in the [3]-symmetric model space,
one misses the dominant [3](04) component, and consequently
one obtains a broad solution with a smaller binding energy.
This is the situation which happens
in Refs.\,\cite{TB03} and \cite{DE03}.

In order to formulate a precise $3\alpha$ OCM equation
with the almost redundant Faddeev components,
we write $\phi^{[3]}_\tau$ with $\tau \sim -1$ as $\Psi_0$,
and define a new projection
operator $\widetilde{P}=|\Psi_0 \rangle \langle \Psi_0|+P$
with $P=\sum_{\lambda=0}|\Psi_\lambda \rangle
\langle \Psi_\lambda |$.
Here $\Psi_\lambda$ with $\lambda=0$ are the [3]-symmetric
Pauli-allowed $3\alpha$ states and $P|\Psi_0\rangle=0$ is
satisfied. 
The $3\alpha$ OCM equation solved in the present formalism is
\begin{eqnarray}
\widetilde{P} [\,E-H_0-\sum_\alpha V^{\rm BFW}_\alpha\,]
\widetilde{P} \Psi=0\ .
\label{eq14}
\end{eqnarray}
(On the other hand, the original equation
with $\widetilde{P} \rightarrow P$ is solved
in Refs.\,\cite{TB03} and \cite{DE03}
in the method of orthogonalizing pseudo-potentials.)
This equation is equivalent with the following two equations:
\begin{subequations}
\label{eq15}
\begin{eqnarray}
& & \langle \Psi_0|\,E-H_0-\sum_\alpha V^{\rm BFW}_\alpha \,|
\widetilde{P} \Psi \rangle=0\ ,\label{eq15a} \\
& & P\,[\,E-H_0-\sum_\alpha V^{\rm BFW}_\alpha\,]
| \widetilde{P} \Psi\rangle=0\ .\label{eq15b}
\end{eqnarray}
\end{subequations}
From \eq{eq15a}, we find 
\begin{eqnarray}
\langle \Psi_0|H|P \Psi \rangle
=(E-E_0) \langle \Psi_0|\Psi\rangle\ ,
\label{eq16}
\end{eqnarray}
where $H=H_0+\sum_\alpha V^{\rm BFW}_\alpha$
and $E_0=\langle \Psi_0|H|\Psi_0\rangle$.
By multiplying \eq{eq15b} by $\langle \Psi|$ from the left
and using \eq{eq16}, we immediately obtain
\begin{eqnarray}
E\,\langle P \Psi|P \Psi \rangle-\langle P \Psi|H|P \Psi \rangle
={| \langle \Psi_0|H|P \Psi \rangle |^2 \over E-E_0}\ .
\label{eq17}
\end{eqnarray}
If $|\Psi_0\rangle$ is an approximate eigenstate of 
the full Hamiltonian $H$,
the coupling term in \eq{eq16} is almost zero.
In this case, we find two solutions for $E$ from
a simple illustration of the graph for \eq{eq17} with
respect to $E$; namely,
one is the $\Psi_0$-dominant ground state with $E\sim E_0$ and
the other the excited state
with $E\sim \langle P \Psi|H|P \Psi \rangle
/\langle P \Psi|P \Psi \rangle$ and a small
admixture of the $\Psi_0$ component.

It is also possible to derive a Faddeev equation
which is completely equivalent to \eq{eq14}.
We assume $|u\rangle$ the bound-state
solution of $V^{\rm BFW}$ with the energy
eigenvalue $\varepsilon_B$. For ${\cal V}_\alpha (E)=E-H_0
-\Lambda_\alpha (E-H_0-V^{\rm BFW}_\alpha) \Lambda_\alpha$ 
with $\Lambda_\alpha=1-|u_\alpha \rangle \langle u_\alpha|$,
one can prove
\begin{eqnarray}
{\cal V}_\alpha (E)-V^{\rm BFW}_\alpha
=|u_\alpha \rangle \langle u_\alpha|(E-h_{\bar \alpha}
-\varepsilon_B)|u_\alpha \rangle \langle u_\alpha|,
\label{eq18}
\end{eqnarray}
where $h_{\bar \alpha}$ is the kinetic-energy operator
of the third $\alpha$-particle.
Owing to this relationship,
we can replace $V^{\rm BFW}_\alpha$ in \eq{eq15b} by
${\cal V}_\alpha (E)$. Following the same procedure as
developed in Ref.\,\cite{TRGM} for $\widetilde{P}\Psi
=\varphi_\alpha+\varphi_\beta+\varphi_\gamma$, we can derive
\begin{eqnarray}
\varphi=G_0 \widetilde{T} S \varphi
+\sum_{\tau \sim -1} |u\widetilde{f}^\tau\rangle
\langle \Psi_0 | \widetilde{P} \Psi \rangle\ ,
\label{eq19}
\end{eqnarray}
where $\widetilde{f}^\tau=\langle u|\Psi_0\rangle
=\sqrt{(1+\tau)/3}\,f^\tau$.
In the intermediate step, we also find
\begin{eqnarray}
& & [\,E-H_0-\sum_\alpha {\cal V}_\alpha (E)\,]
\widetilde{P}\Psi \nonumber \\
& & =-\sum_\alpha | u_\alpha \rangle
\langle u_\alpha |\,E-H_0\,|
\varphi_\beta+\varphi_\gamma \rangle\ .
\label{eq20}
\end{eqnarray}
We multiply this equation by $\langle \Psi_0|$ from the left,
and subtract the resultant equation from \eq{eq15a}.
Then the symmetry of the matrix elements yields
$\langle \Psi_0 |({\cal V}_\alpha (E)-V^{\rm BFW}_\alpha)
|\widetilde{P}\Psi \rangle
=\langle u_\alpha \widetilde{f}^\tau_\alpha |\,E-H_0\,|
\varphi_\beta+\varphi_\gamma \rangle$
for each $\alpha$ and $\tau \sim -1$,
or (from \eq{eq18} and restoring $V^{\rm BFW}$)
\begin{eqnarray}
\langle \Psi_0 | \widetilde{P}\Psi\rangle 
& = & {\langle u \widetilde{f}^\tau |\,(E-H_0)S\,|
\varphi \rangle \over
\langle u \widetilde{f}^\tau|\,E-H_0-V^{\rm BFW}\,|
u \widetilde{f}^\tau \rangle}\ .
\label{eq21}
\end{eqnarray}
Using this relationship in \eq{eq19}, we eventually obtain
a new type of Faddeev equation
\begin{eqnarray}
\varphi & = & G_0 \widetilde{T} S \varphi
+\sum_{\tau \sim -1} |u f^\tau\rangle
{1 \over \langle u f^\tau|E-H_0-V^{\rm BFW}|u f^\tau\rangle}
\nonumber \\
& & \times \langle u f^\tau|\,(E-H_0)S\,|\varphi\rangle\ .
\label{eq22}
\end{eqnarray}
The solutions of this equation are given in Ref.\,\cite{ocm03},
together with the results of the direct variational calculations
of \eq{eq14}, using the translationally invariant h.o. basis.

In summary, we probably cannot obtain the [3](04)-dominant
compact ground state without a small admixture of the
redundant components, which is related with the model space
character that no exact [21]-symmetric solution exists in the
model space $|uf^\tau \rangle$, when the BFW bound-state solution
is used for $|u\rangle$. 
If one insists mathematical rigorousness that the forbidden
components should be completely eliminated from the exact solution,
we have to say that there is no compact bound state possible
in the $3\alpha$ problem for the BFW $\alpha \alpha$ potential.
We, however, keep in mind that the orthogonality condition
model is just a model which takes into account
the major roles of the Pauli principle among clusters.
From the microscopic viewpoint based on the resonating-group
method, a small admixture of the redundant components is
easily swept away by the effect of antisymmetrization.
It is our opinion that the description of the physical ground
state of the $3\alpha$ system with the compact shell-model like
structure is far more important than the strict demand to eliminate
the redundant components.

\bigskip

\begin{acknowledgments}

We wish to thank D. Baye, P. Descouvemont and E. M. Tursunov
for many useful discussions.
This work was supported by Grants-in-Aid for Scientific
Research (C) from Japan Society for the Promotion
of Science (JSPS) (Nos. 15540270, 15540284).
One of the authors (Y. F.) also thanks FNRS foundation
of Belgium for making his visit to Free University
of Brussels possible during the summer, 2002.

\end{acknowledgments}

\end{document}